\documentclass[pre,preprintnumbers,amsmath,amssymb]{revtex4}
\usepackage{amssymb,epsf}
\usepackage{latexsym}                                  %Text revised on 26 Sep 2011%
\usepackage{graphicx}
\usepackage{float}
\begin{document}
\title{Spherically symmetric solutions in a FRW background}
\author{H. Moradpour$^1$\footnote{h.moradpour@riaam.ac.ir} and N. Riazi$^2$\footnote{n$\_$riazi@sbu.ac.ir}}
\address{$1$ Research Institute for Astronomy and Astrophysics of Maragha (RIAAM), P.O. Box 55134-441, Maragha, Iran,\\
$^2$ Physics Department, Shahid Beheshti University,
Evin, Tehran 19839, Iran.}

\begin{abstract}
We impose perfect fluid concept along with slow expansion
approximation to derive new solutions which, considering non-static
spherically symmetric metrics, can be treated as Black Holes. We
will refer to these solutions as Quasi Black Holes. Mathematical and
physical features such as Killing vectors, singularities, and mass
have been studied. Their horizons and thermodynamic properties have
also been investigated. In addition, relationship with other related
works (including mcVittie's) are described.
\end{abstract}
\keywords{Conformal transformation; perfect fluid; dynamical black
holes; black hole thermodynamics.}
%\ccode{PACS Nos.: 04.20.-q; 04.70.-s.}
\maketitle
\bigskip
\section{Introduction \label{Introduction}}
The Universe expansion can be modeled by the so called FRW metric
\begin{eqnarray}\label{FRW}
ds^2=-dt^2+a(t)^2[\frac{dr^2}{(1-kr^2)}+r^2d\theta^2+r^2sin(\theta)^2d\phi^2],
\end{eqnarray}
where $k=0, +1, -1$ are curvature scalars which represent the flat,
closed and open universes, respectively. The WMAP data confirms a
flat ($k=0$) universe \cite{Roos}. $a(t)$ is the scale factor and
for a background which is filled by a perfect fluid with equation of
state $p=\omega \rho$, there are three classes of expanding
solutions. These three solutions are
\begin{eqnarray}\label{Scale factor1}
a(t)=a_0 t^{\frac{2}{3(\omega + 1)}}
\end{eqnarray}
for $ \omega\neq 0$ when $-1<\omega$ and ,
\begin{eqnarray}\label{Scale factor2}
a(t)=a_0 e^{Ht}
\end{eqnarray}
for $\omega=-1$ (dark energy), and for the Phantom regime
($\omega<-1$) is
\begin{eqnarray}\label{Scale factor3}
a(t)=a_0(t_0-t)^{\frac{2}{3(\omega + 1)}},
\end{eqnarray}
where $t_0$ is the big rip singularity time and will be available,
if the universe is in the phantom regime.

In Eq.~(\ref{Scale factor2}), $H(\equiv\frac{\dot{a}(t)}{a(t)})$ is
the Hubble parameter and the current estimates are
$H=73^{+4}_{-3}kms^{-1}Mpc^{-1}$ \cite{Roos}.

Note that, at the end of the Phantom regime, everything will
decompose into its fundamental constituents \cite{Mukh}. In
addition, this spacetime can be classified as a subgroup of the
Godel-type spacetime with $\sigma=m=0$ and $k^{\prime}=1$
\cite{godel}.

A signal which was emitted at the time $t_0$ by a co-moving source
and absorbed by a co-moving observer at a later time $t$ is affected
by a redshift ($z$) as
\begin{eqnarray}
1+z=\frac{a(t)}{a(t_0)}.
\end{eqnarray}
The apparent horizon as a marginally trapped surface, is defined as
\cite{SWR}
\begin{eqnarray}\label{aph1}
g^{\mu \nu}\partial_{\mu}\xi \partial_{\nu}\xi=0,
\end{eqnarray}
which for the physical radius of $\xi=a(t)r$, the solution will be:
\begin{eqnarray}\label{aph2}
\xi=\frac{1}{\sqrt{H^2+\frac{k}{a(t)^2}}}.
\end{eqnarray}
The surface gravity of the apparent horizon can be evaluated by:
\begin{eqnarray}\label{sg1}
\kappa=\frac{1}{2\sqrt{-h}}\partial_a(\sqrt{-h}h^{ab}\partial_b
\xi).
\end{eqnarray}

Where the two dimensional induced metric is
$h_{ab}=diag(-1,\frac{a(t)}{(1-kr^2)})$. It was shown that the first
law of thermodynamics is satisfied on the apparent horizon
\cite{S0,S1,S2,S3}. The special case of $\omega=-1$ is called the
dark energy, and by a suitable change of variables one can rewrite
this case in the static form \cite{Poisson}:
\begin{eqnarray}\label{static dark}
ds^2=-(1-H^2r^2)dt^2+\frac{dr^2}{(1-H^2r^2)}+r^2d\Omega^2.
\end{eqnarray}
This metric belongs to a more general class of spherically
symmetric, static metrics. For these class of spherically symmetric
static metrics, the line element can be written in the form of:
\begin{eqnarray}\label{SSM}
ds^2=-f(r)dt^2+\frac{dr^2}{f(r)}+r^2d\Omega^2,
\end{eqnarray}
where the general form of $f(r)$ is:
\begin{eqnarray}
f(r)=1-2\frac{m}{r}+\frac{Q^2}{r^2}-H^2r^2.
\end{eqnarray}
In the above expression, $m$ and $Q$ represent mass and charge,
respectively. For this metric, one can evaluate redshift:
\begin{eqnarray}
1+z=(\frac{1-2\frac{m}{r}+\frac{Q^2}{r^2}-H^2r^2}
{1-2\frac{m}{r_0}+\frac{Q^2}{r_0^2}-H^2r_0^2})^\frac{1}{2}.
\end{eqnarray}
Where, $r_0$ and $r$ are radial coordinates at the emission and the
absorption points. For the horizons, the radius and the surface
gravity can be found using equations
\begin{eqnarray}\label{SG}
g_{tt}&=&f(r)=0 \longrightarrow r_h \\ \nonumber \kappa
&=&\frac{f^\prime(r)}{2}|_{r_h},
\end{eqnarray}
where $(^\prime)$ denotes derivative with respect to the coordinate
$r$ \cite{Poisson}. From the thermodynamic laws of Black Holes (BHs)
we know
\begin{eqnarray}\label{Temp2}
T=\frac{\kappa}{2\pi},
\end{eqnarray}
which $T$ is the temperature on the horizon \cite{Poisson}. Validity
of the first law of the thermodynamics on the static horizons for
the static spherically symmetric spacetime has been shown
\cite{Cai1,Padm1}.

The BHs with the FRW dynamic background has motivated many
investigations. The first approach, which is named Swiss Cheese,
includes efforts in order to find the effects of the expansion of
the Universe on the gravitational field of the stars \cite{P1},
introduced originally by Einstein and Straus $(1945)$ \cite{ES}. In
these models, authors tried to join the Schwarzschild metric to the
FRW metric by satisfying the junction conditions on the boundary,
which is an expanding timelike hypersurface. The inner spacetime is
described by the Schwarzschild metric, while the FRW metric explains
the outer spacetime. These models don't contain dynamical BHs,
Because the inner spacetime is in the Schwarzschild coordinate,
hence, is static \cite{saida}. In addition, the Swiss Cheese models
can be classified as a subclass of inhomogeneous
Lemabitre-Tolman-Bondi models \cite{MD1,CLure}.

Looking for dynamical BHs, some authors used the conformal
transformation of the Schwarzschild BH, where the conformal factor
is the scale factor of the famous FRW model. Originally, Thakurta
$(1981)$ have used this technique and obtained a dynamical version
of the Schwarzschild BH \cite{Thak}. Since the Thakurta spacetime is
a conformal transformation of the Schwarzschild metric, it is now
accepted that its redshift radii points to the co-moving radii of
the event horizon of BH \cite{MD1,MR,RMS}. By considering asymptotic
behavior of the gravitational lagrangian (Ricci scalar), one can
classify the Thakurta BH and its extension to the charged BH into
the same class of solutions \cite{MR,RMS}. The Thakurta spacetime
sustains an inward flow, which leads to an increase in the mass of
BH \cite{MR,RMS,Gao1}. This ingoing flow comes from the
back-reaction effect and can be neglected in a low density
background \cite{Gao1}. In fact, for the low density background, the
mass will be decreased in the Phantom regime \cite{Bab}. Also, the
radius of event horizon increases with the scale factor when its
temperature decreases by the inverse of scale factor \cite{MR,RMS}.

Using the Eddington-Finkelstein form of the Schwarzschild metric and
the conformal transformation, Sultana and Dyer $(2005)$ have
constructed their metric and studied its properties \cite{SD}. In
addition, unlike the Thakurta spacetime, the curvature scalars do
not diverge at the redshift singularity radii (event horizon) of the
Sultana and Dyer spacetimes. Since the Sultana and Dyer spacetimes
is conformal transformation of the Schwarzschild metric, it is now
accepted that the Sultana and Dyer spacetimes include dynamic BHs
\cite{MD1}. Various examples can be found in \cite{MD1,MD2,FJ,MN}.
Among these conformal BHs, only the solutions by
M$^{\textmd{c}}$Clure et al. and Thakurta can satisfy the energy
conditions \cite{RMS,MD1}. Static charged BHs which are confined
into the FRW spacetime and the dynamic, charged BHs were studied in
\cite{O1,O2,O3,O4,O5,O6,O7,O8}. The Brane solutions can be found in
\cite{BS1,BS2,BS3}.

In another approach, mcVittie found new solutions including
contracting BHs in the coordinates co-moving with the universe's
expansion \cite{mcvittie}. Its generalization to the arbitrary
dimensions and to the charged BHs can be found in \cite{Gao0,Gao}.
In these solutions, it is easy to check that the curvature scalars
diverge at the redshift singularities. In this approach, authors
have used the isotropic form of the FRW metric along as the perfect
fluid concept and could find their solutions which can contain BHs
\cite{Far}. The mass and the charge of their BHs seem to be
decreased with the scale factor. Also, it seems that the redshift
singularities does not point to a dynamic event horizon
\cite{nol1,nol2,SUS,fri}. Unlike the Swiss Cheese models, the energy
conditions are violated by these solutions \cite{MD1}. These
solutions can be considered as Models for cosmological
inhomogeneities \cite{CLure}.

This paper is organized as follows: in the next section, we consider
the conformal transformation of a non-static spherically symmetric
metric, where conformal factor has only time dependency. In
addition, we derive the general possible form of metric by using
perfect fluid concept. In section $3$, slow time varying
approximation is used in order to find the physical meaning of the
parameters of metric. In continue, the mcVittie like solution and
its thermodynamic properties are addressed. In section $4$, we
generalized our debates to the charged spacetime, when the effects
of the dark energy are considerable. In section $5$, we summarize
and conclude the results.
\section{Metric, general properties and basic assumptions}
Let us begin with this metric:
\begin{eqnarray}
ds^2=a(\tau)^2[-f(\tau,r)d\tau^2+\frac{dr^2}{(1-kr^2)f(\tau,r)}+r^2d\theta^2+r^2sin(\theta)^2d\phi^2].
\end{eqnarray}
Where $a(\tau)$ is the arbitrary function of time coordinate $\tau$.
This metric has three Killing vectors
\begin{eqnarray}\label{symm}
\partial_{\phi},\ \ \sin\phi \ \partial_{\theta}+\cot\theta \ \cos\phi \
 \partial_{\phi}\ \ \textmd{and}\ \ \cos\phi \ \partial_\theta - \cot\theta \ \sin\phi \ \partial_\phi .
\end{eqnarray}
Now, if we define new time coordinate as
\begin{eqnarray}
\tau \rightarrow t=\int a(\tau)d\tau,
\end{eqnarray}
we will get
\begin{eqnarray}\label{Metric1}
ds^2=-f(t,r)dt^2+a(t)^2[\frac{dr^2}{(1-kr^2)f(t,r)}+r^2d\theta^2+r^2sin(\theta)^2d\phi^2],
\end{eqnarray}
which possesses symmetries like as Eq.~(\ref{symm}). From now, it is
assumed that $a(t)$ is the cosmic scale factor similar to the FRW's.
For $f(t,r)=1$, Eq.~(\ref{Metric1}) is reduced to the FRW metric
(\ref{FRW}). Also, conformal BHs can be achieved by choosing
$f(t,r)=f(r)$ where, the general form of $f(r)$ is \cite{MR}:
\begin{eqnarray}
f(r)=1-\frac{2m}{r}+\frac{Q^2}{r^2}-\frac{\Lambda r^2}{3}.
\end{eqnarray}
Therefore, conformal BHs can be classified as a special subclass of
metric (\ref{Metric1}). $n_{\alpha}=\delta^r_{\alpha}$ is normal to
the hypersurface $r=const$ and yields
\begin{eqnarray}\label{nhs}
n_{\alpha}n^{\alpha}=g^{rr}=\frac{(1-kr^2)f(t,r)}{a(t)^2},
\end{eqnarray}
which is timelike when $(1-kr^2)f(t,r)<0$, null for
$(1-kr^2)f(t,r)=0$ and spacelike if we have $(1-kr^2)f(t,r)>0$. For
an emitted signal at the coordinates $t_0$ and $r_0$, when it is
absorbed at coordinates $t$ and $r$ simple calculations lead to
\begin{eqnarray}\label{redshift}
1+z=\frac{\lambda}{\lambda_0}=\frac{a(t)}{a(t_0)}(\frac{f(t,r)}{f(t_0,r_0)})^{\frac{1}{2}},
\end{eqnarray}
as induced redshift due to the universe expansion and factor
$f(t,r)$. Redshift will diverge when $f(t_0 , r_0)$ goes to zero or
$1+z\longrightarrow \infty$. This divergence as the signal of
singularity is independent of the curvature scalar ($k$), unlike the
Mcvittie's solution and its various generalizations \cite{Gao0,Gao},
which shows that our solutions are compatible with the FRW
background. As a desired expectation, it is obvious that the FRW
result is covered when $f(t_0 , r_0)=f(t,r)=1$. The only
non-diagonal term of the Einstein tensor is
\begin{eqnarray}\label{off diagonal}
G^{t
r}=-\frac{1-kr^2}{f(t,r)a(t)^3r}(a(t)\dot{f}(t,r)-{f}^\prime(t,r)\dot{a}(t)r),
\end{eqnarray}
which $(\dot{})$ and $(^\prime)$ are derivatives with respect to
time and radius, respectively. Using $\frac{\partial f}{\partial
t}=\dot{a}\frac{\partial f}{\partial a}$, one gets
\begin{eqnarray}\label{off diagonal2}
G^{t
r}=-\frac{(1-kr^2)\dot{a}(t)}{f(a(t),r)a(t)^3r}(a(t)\tilde{f}(a(t),r)-{f}^\prime(a(t),r)r),
\end{eqnarray}
where $\tilde{f}(a(t),r)=\frac{\partial f}{\partial a}$. In order to
get perfect fluid solutions, we impose condition $G^{tr}=0$ and
reach to
\begin{eqnarray}\label{f}
f(t,r)=f(a(t)r)=\sum_n b_n (a(t)r)^n.
\end{eqnarray}
Although Eq.~(\ref{f}) includes numerous terms, but the slow
expansion approximation helps us to attribute physical meaning to
the certain coefficients $b_n$. Since $G_{tr}=0$, we should stress
that here that there is no redial flow and thus, the backreaction
effect is zero \cite{RMS,Gao1}, which means that there is no energy
accretion in these solutions \cite{fj}. Finally and briefly, we see
that the perfect fluid concept is in line with the no energy
accretion condition. The only answer which is independent of the
rate of expansion can be obtained by condition $b_n=\delta_{n0}$
which is yielding the FRW solution.
\section{mcVittie like solution in the FRW background}
The mcVittie's solution in the flat FRW background can be written as
\cite {MD1}
\begin{eqnarray}\label{mc1}
ds^2=-(\frac{1-\frac{M}{2a(t)\tilde{r}}}{1+\frac{M}{2a(t)\tilde{r}}})^2dt^2+
a(t)^2 (1+\frac{M}{2a(t)\tilde{r}})^4[d\tilde{r}^2+
\tilde{r}^2d\Omega^2].
\end{eqnarray}
This metric possess symmetries same as metric~(\ref{Metric1}).
$\tilde{r}$ is isotropic radius defined by:
\begin{eqnarray}\label{r1}
r=\tilde{r}(1+\frac{M}{2 \tilde{r}})^2.
\end{eqnarray}
There is a redshift singularity at radii
$\tilde{r}_h=\frac{M}{2a(t)}$ which yields the radius
$r_h=\frac{M}{2a(t)}(1+a(t))^2$ \cite{Fara}. In addition,
$\tilde{r}_h$ is a spacelike hypersurface, and can not point to an
event horizon \cite{fj}.

Consider $f(a(t)r)=1-\frac{2b_{-1}}{a(t)r}$. This assumption
satisfies condition~(\ref{f}) and leads to
\begin{eqnarray}\label{SCHWM}
ds^2=-(1-\frac{2b_{-1}}{a(t)r})dt^2+a(t)^2
[\frac{dr^2}{(1-kr^2)(1-\frac{2b_{-1}}{a(t)r})}+r^2d\Omega^2].
\end{eqnarray}
For $b_{-1}\neq0$, this metric will converge to the FRW metric when
$r\longrightarrow \infty$. The Schwarzschild metric is obtainable by
putting $a(t)=1$, $b_{-1}=M$ and $k=0$. Metric suffers from three
singularities at $a(t)=0$ (big bang), $r=0$ and
\begin{eqnarray}\label{SCHR}
f(a(t)r)=0\Rightarrow a(t)r_h=2b_{-1}.
\end{eqnarray}
Third singularity exists if $b_{-1}>0$. In this manner,
Eq.~(\ref{redshift}) will diverge at $r_0=r_h$. In addition and in
contrast to the Gao's solutions, the radii of the redshift
singularity $(r_h)$ in our solutions is independent of the
background curvature $(k)$, while for the flat case our radius is
compatible with the previous works \cite{mcvittie,Gao,MD1}. Also,
metric changes its sign at $r=r_h$ just the same as the
schwarzschild spacetime. In addition, curvature scalars diverge at
this radius as well as the mcVittie spacetime. Accordingly, this
singularity point to a naked singularity which can be considered as
alternatives for BHs \cite{virb,virb1}. In continue, we will point
to the some physical and mathematical properties of this singularity
which has the same behaviors as event horizon if one considers slow
expansion approximation. The surface area integration at this radius
leads to
\begin{eqnarray}\label{surface area1}
A=\int\sqrt{\sigma}d\theta d\phi=4\pi r^2_h a(t)^2=16\pi
(b_{-1})^{2}.
\end{eqnarray}
The main questions that arise here are: what is the nature of
$b_{-1}$? and can we better clarify the meaning of $r_h$? For these
purposes, we consider the slow expansion approximation $(a(t)\approx
c)$, define new coordinate $\eta=cr$ and get
\begin{eqnarray}\label{apmetric}
ds^2\approx-(1-\frac{2b_{-1}}{\eta})dt^2+\frac{d\eta^2}{(1-k^{\prime}\eta^2)
(1-\frac{2b_{-1}}{\eta})}+\eta^2d\theta^2+\eta^2sin(\theta)^2d\phi^2,
\end{eqnarray}
where $k^{\prime}=\frac{k}{c^2}$. In these new coordinates,
$(t,\eta,\theta,\phi)$, and from Eq.~(\ref{nhs}) it is apparent that
for $b_{-1}>0$, hypersurface with equation $\eta=\eta_h=2b_{-1}$ is
a null hypersurface. When our approximation is broken, then $\eta_h$
may not be actually a null hypersurface, despite its resemblance to
that. We call this null hypersurface a quasi event horizon which is
signalling us an object like a BH and we refer to that as a quasi
BH. From now, we assume $b_{-1}>0$, the reason of this option will
be more clear later, when we debate mass. Therefore by the slow
expansion approximation, $r_h$ ($=\frac{2b_{-1}}{c}$) plays the role
of the co-moving radius of event horizon and it is decreased with
time. In order to find an answer to the first question about the
physical meaning of $b_{-1}$, we use Komar mass:
\begin{eqnarray}\label{mass1}
M=\frac{1}{4\pi}\int_S  n^{\alpha} \sigma_{\beta}
\triangledown_\alpha \xi^\beta_t dA,
\end{eqnarray}
where $\xi^\beta_t$ is the timelike Killing vector of spacetime.
Since the Komar mass is only definable for the stationary and
asymptotically flat spacetimes \cite{W1}, one should consider the
flat case ($k=0$) and then by bearing the spirit of the stationary
limit in mind (the slow expansion approximation) tries to evaluate
Eq.~(\ref{mass1}).

Consider $n_{\alpha}=\sqrt{1-\frac{2b_{-1}}{a(t)r}}\delta^t_\alpha$
and
$\sigma_\beta=\frac{a(t)}{\sqrt{1-\frac{2b_{-1}}{a(t)r}}}\delta^r_\beta$
as the unit timelike and unit spacelike four-vectors, respectively.
Now using Eq.~(\ref{mass1}) and bearing the spirit of the slow
expansion approximation in mind, one gets
\begin{eqnarray}\label{mass}
M=\frac{1}{4\pi}\int_S n^{\alpha} \sigma_{\beta}
\Gamma^{\beta}_{\alpha t}dA=b_{-1},
\end{eqnarray}
which is compatible with the no energy accretion condition
($G_{tr}=0$). In addition, we will find the same result as
Eq.~(\ref{mass}), if we considered the flat case ($k=0$) of
metric~(\ref{apmetric}) and use
$n_{\alpha}=\sqrt{1-\frac{2b_{-1}}{\eta}}\delta^t_\alpha$ and
$\sigma_\beta=\frac{1}{\sqrt{1-\frac{2b_{-1}}{\eta}}}\delta^\eta_\beta$.
Since the integrand is independent of the scale factor ($a(t)$), the
slow expansion approximation does not change the result of integral.
But, the accessibility of the slow expansion approximation is
necessary if one wants to evaluate the Komar mass for dynamical
spacetimes \cite{W1}. Indeed, this situation is the same as what we
have in the quasi-equilibrium thermodynamical systems, where the
accessibility of the quasi-equilibrium condition lets us use the
equilibrium formulation for the vast thermodynamical systems
\cite{callen}. It is obvious that for avoiding negative mass, we
should have $b_{-1}>0$. Relation to the Komar mass of the mcVittie's
solution can be written as \cite{MD1,Gao}
\begin{eqnarray}\label{komar}
M_{mcVittie}=\frac{M}{a(t)}.
\end{eqnarray}
In addition, some studies show that the Komar mass is just a metric
parameter in the mcVittie spacetime \cite{nol1,nol2,fj}. Indeed,
Hawking-Hayward quasi-local mass satisfies $\dot{M}=0$, which is
compatible with $G_{r t}=0$ and indicates that there is no redial
flow and thus the backreaction effect, in the mcVittie's solution
\cite{RMS,Gao1,Bab,fj}. In order to clarify the mass notion in the
mcVittie spacetime, we consider the slow expansion approximation of
the mcVittie spacetime which yields
\begin{eqnarray}\label{mse}
ds^2 \approx -(\frac{1-\frac{M}{2\eta}}{1+\frac{M}{2\eta}})^2dt^2+
(1+\frac{M}{2\eta})^4[d\eta^2+ \eta^2d\Omega^2].
\end{eqnarray}
This metric is signalling us that the $M$ may play the role of the
mass in the mcVittie spacetime. In addition, by defining new radii
$R$ as
\begin{eqnarray}
R(t,r)=a(t)\tilde{r}(1+\frac{M}{2\tilde{r}})^2,
\end{eqnarray}
one can rewrite the mcVittie spacetime in the form of
\begin{eqnarray}\label{mn}
ds^2=-(1-\frac{2M}{R}-H^2R^2)dt^2-\frac{2HR}{\sqrt{1-\frac{2M}{R}}}dtdR+
\frac{dR^2}{1-\frac{2M}{R}}+R^2d\Omega^2,
\end{eqnarray}
where $H=\frac{\dot{a}}{a}$ \cite{ff}. This form of the mcVittie
spacetime indicates these facts that the Komar mass is a metric
parameter and $M$ is the physical mass in this spacetime \cite{ff}.
Finally, we see that the results of the slow expansion approximation
(Eq.~(\ref{mse})) and Eq.~(\ref{mn}) are in line with the result of
the study of the Hawking-Hayward quasi-local mass in the mcVittie
spacetime \cite{nol1,nol2,fj,ff}. For the flat case ($k=0$) of our
spacetime (Eq.~(\ref{SCHWM})), by considering Eq.~(\ref{komar}) and
following the slow expansion approximation, we reach at
\begin{eqnarray}\label{nm1}
 ds^2\approx-(1-\frac{2M}{\eta})dt^2+\frac{d\eta^2}{(1-\frac{2M}{\eta})}
 +\eta^2d\theta^2+\eta^2sin(\theta)^2d\phi^2.
\end{eqnarray}
Also, if we define new radius $R$ as
\begin{eqnarray}
r=\frac{R}{a}(1+\frac{M}{2R})^2,
\end{eqnarray}
we obtain
\begin{eqnarray}\label{nm}
ds^2&=&
-(\frac{(1-\frac{M}{2R})^2}{(1+\frac{M}{2R})^2}-\frac{R^2H^2(1+\frac{M}{2R})^6}
{(1-\frac{M}{2R})^2})dt^2-
\frac{2RH(1+\frac{M}{2R})^5}{(1-\frac{M}{2R})}dtdR\\ \nonumber
&+&(1+\frac{M}{2R})^4[dR^2+R^2d\Omega^2].
\end{eqnarray}
Both of the equations (\ref{nm1}) and (\ref{nm}) as well as the no
energy accretion condition suggest that, unlike the mcVittie's
spacetime, the Komar mass may play the role of the mass in our
solution. From Eq.~(\ref{nm}) it is apparent that $R=\frac{M}{2}$
points to the spacelike hypersurface where, in the
metric~(\ref{mn}), $R=2M$ points to the null hypersurface. In the
next subsection and when we debate thermodynamics, we will derive
the same result for the mass notion in our spacetime.Only in the
$a(t)=1$ limit (the Schwarzschild limit), Eqs.~(\ref{nm})
and~(\ref{mc1}) will be compatible which shows that our spacetime is
different with the mcVittie's. Let us note that the obtained metric
(Eq.~(\ref{nm})) is consistent with Eq.~(\ref{mn}), provided we take
$M=0$ (the FRW limit).
\subsection*{Horizons, energy and thermodynamics}
There is an apparent horizon in accordance with the FRW background
which can be evaluated from Eq.~(\ref{aph1}):
\begin{eqnarray}
(1-kr_{ap}^2)(1-\frac{2M}{a(t)r_{ap}})^2-r_{ap}^2\dot{a}(t)^2=0.
\end{eqnarray}
This equation covers the FRW results in the limit of
$M\longrightarrow0$ ( see Eq.~(\ref{aph2})). In addition, one can
get the Schwarzschild radius by considering $\dot{a}(t)=0$, which
supports our previous definition for $b_{-1}$. Calculations for the
flat case yield four solutions. The only solution which is in full
agreement with the limiting situation of the FRW metric (in the
limit of zero $M$) is
\begin{eqnarray}\label{hcr}
r_{ap}=\frac{1+\sqrt{1-8HM}}{2\dot{a}}.
\end{eqnarray}
Therefore, the physical radius of apparent horizon
$(\xi_{ap}=a(t)r_{ap})$ is
\begin{eqnarray}
\xi_{ap}=\frac{1+\sqrt{1-8HM}}{2H},
\end{eqnarray}
which is similar to the conformal BHs \cite{RMS}. It is obvious that
in the limit of $M\longrightarrow0$, the radius for the apparent
horizon of the flat FRW is recovered. For the surface gravity of
apparent horizon, one can use Eq.~(\ref{sg1}) and gets:
\begin{eqnarray}\label{fsg}
\kappa=\frac{\kappa_{FRW}}{(1-\frac{2M}{a(t)r_{ap}})^2}+\frac{M}{a(t)^2}
[\frac{1}{r_{ap}^2}+\frac{1}{(1-\frac{2M}{a(t)r_{ap}})^2}(\ddot{a}(t)+
2\frac{\dot{a}(t)^2}{a(t)})],
\end{eqnarray}
where
$h^{ab}=diag(-\frac{1}{1-\frac{2M}{a(t)r}},\frac{1-\frac{2M}{a(t)r}}{a(t)^2})$,
$r_{ap}$ is the apparent horizon co-moving radius~(\ref{hcr}) and
$\kappa_{FRW}$ is the surface gravity of the flat FRW manifold
\begin{eqnarray}
\kappa_{FRW}=-\frac{\dot{a}(t)^2+a(t)\ddot{a}(t)}{2a(t)\dot{a}(t)}.
\end{eqnarray}

The schwarzschild limit ($\kappa=\frac{1}{4M}$) is obtainable by
inserting $a(t)=1$ in Eq.~(\ref{fsg}). In the limiting case
$M\longrightarrow0$, Eq.~(\ref{fsg}) is reduced to the surface
gravity of the flat FRW spacetime, as a desired result. The
Misner-Sharp mass inside radius $\xi$ for this spherically symmetric
spacetime is defined as \cite{Ms}:
\begin{eqnarray}\label{MS}
M_{MS}=\frac{\xi}{2}(1-h^{ab}\partial_a \xi \partial_b \xi).
\end{eqnarray}
Because this definition does not yield true results in some theories
such as the Brans-Dicke and scalar-tensor gravities, we are pointing
to the Gong-Wang definition of mass \cite{Gong}:
\begin{eqnarray}\label{GW}
M_{GW}=\frac{\xi}{2}(1+h^{ab}\partial_a \xi \partial_b \xi).
\end{eqnarray}
It is apparent that, for the apparent horizon, Eqs.~(\ref{MS}) and
(\ref{GW}) yield the same result as
$M_{GW}=M_{MS}=\frac{\xi_{ap}}{2}$. In the limit of
$M\longrightarrow0$, the FRW's results are recovered and we reach to
$M_{GW}=M_{MS}=\rho V$ as a desired result \cite{Cai1}. Using
Eqs.~(\ref{MS}) and~(\ref{GW}) and taking the slow expansion
approximation into account, we reach to $M_{GW}=M_{MS}\simeq M$ as
the mass of quasi BH. Also, this result supports our previous guess
about the Komar mass as the physical mass in our solution, and is in
line with the result of Eqs.~(\ref{nm1}) and~(\ref{nm}). For the
Mcvittie metric, Eqs.~(\ref{MS}) and (\ref{GW}) yield
$M_{GW}=M_{MS}\simeq\frac{M}{4}$ as the confined mass to radius
$\xi_h=a(t)\tilde{r}_h=\frac{M}{2}$. Also,
Eqs.~(\ref{mass}),~(\ref{MS}) and~(\ref{GW}) leads to the same
result in the Schwarzschild's limit ($\mathcal{M}=M_{GW}=M_{MS}=M$).
For the flat background, using metric~(\ref{apmetric}),
Eq.~(\ref{SG}) and inserting results into Eq.~(\ref{Temp2}), one
gets
\begin{eqnarray}\label{Temp1}
T\simeq\frac{1}{8 \pi M},
\end{eqnarray}
for the temperature on the surface of quasi horizon. The same
calculations yield similar results, as the temperature on the
horizon of the Mcvittie's solution. For the conformal Schwarzschild
BH, the same analysis leads to
\begin{eqnarray}
T\simeq\frac{1}{8 \pi a(t)M},
\end{eqnarray}
which shows that the $a(t)M$ plays the role of mass, and is
compatible with the energy accretion in the conformal BHs
\cite{RMS,Gao1,Bab,Fara}. Again, we see that the temperature
analysis can support our expectation from $M$ as the physical mass
in our solutions. For the area of quasi horizon, we have
\begin{eqnarray}\label{surface area}
A=\int\sqrt{\sigma}d\theta d\phi=4\pi a(t)^2 r_h^2=16\pi M^{2}.
\end{eqnarray}
In the mcVittie spacetime, this integral leads to $A=16 \pi M^2$. In
order to vindicate our approximation, we consider $S=\frac{A}{4}$
for the entropy of quasi BH. In continue and from Eq.~(\ref{Temp1}),
we get
\begin{eqnarray}
TdS\simeq dM=dE.
\end{eqnarray}

Whereas, we reach to $TdS\simeq dM\neq dE $ for the mcVittie
spacetime. In the coordinates $(t,\eta,\theta,\phi)$, we should
remind that, unlike the mcVittie spacetime, $E=M_{GW}=M_{MS}\simeq
M$ is valid for quasi BH and the work term can be neglected as the
result of slow expansion approximation ($dW\sim 0$) \cite{Fara}.
Finally and unlike the mcVittie's horizon, we see that $TdS\simeq
dE$ is valid on the quasi event horizon. This result points us to
this fact that the first law of the BH thermodynamics on quasi event
horizon will be satisfied if we use either the Gong-Wang or the
Misner-Sharp definitions for the energy of quasi BH. $TdS\simeq dE$
is valid for the conformal Schwarzschild BH, too \cite{Fara}. For
the flat background, we see that the surface area at redshift
singularity in our spacetime is equal to the mcVittie metric which
is equal to the Schwarzschild metric. In continue and by bearing the
slow expansion approximation in mind, we saw that the temperature on
quasi horizon is like the Schwarzschild spacetime \cite{RMS}. In
addition, we saw that the quality of the validity of the first law
of the BH thermodynamics on quasi event horizon is like the
conformal Schwarzschild BH's and differs from the mcVittie's
solution.

In another approach and for the mcVittie spacetime, if we use the
Hawking-Hayward definition of mass as the total confined energy to
the hypersurface $\tilde{r}=\frac{M}{2a(t)}$, we reach to
\begin{eqnarray}\label{Tempp}
TdS\simeq dM=dE,
\end{eqnarray}
where we have considered the slow expansion approximation. In
addition, Eq.~(\ref{Tempp}) will be not valid, if one uses the Komar
mass~(\ref{komar}). Finally, we saw that the first law of
thermodynamics will be approximately valid in the mcVittie's
solution, if one uses the Hawking-Hayward definition of energy.
Also, none of the Komar, Misner-Sharp and Gong-Wang masses can not
satisfy the first law of thermodynamics on the mcVittie's horizon.
\section{Other Possibilities}
According to what we have said, it is obvious that there are two
other meaningful sentences in expansion~(\ref{f}). The first term is
due to $n=-2$ and points to the charge, where the second term comes
from $n=2$ and it is related to the cosmological constant.
Therefore, the more general form of $f(t,r)$ can be written as:
\begin{eqnarray}\label{tf}
f(t,r)=1-\frac{2M}{a(t)r}+\frac{Q^2}{(a(t)r)^2}-\frac{1}{3}\Lambda
(a(t)r)^2,
\end{eqnarray}
where we have considered the slow expansion approximation and used
these definitions $b_{-2}\equiv Q^2$ and
$b_2\equiv-\frac{1}{3}\Lambda$. Imaginary charge $(b_{-2}<0)$ and
the anti De-Sitter $(\Lambda<0)$ solutions are allowed by this
scheme, but these possibilities are removed by the other parts of
physics. Consider Eq.~(\ref{tf}) when $\Lambda=0$, there are two
horizons located at $r_+=\frac{M+\sqrt{M^2-Q^2}}{a(t)}$ and
$r_-=\frac{M-\sqrt{M^2-Q^2}}{a(t)}$. These radiuses are same as the
Gao's flat case \cite{Gao}. In the low expansion regime ($a(t)\sim
c$), these radiuses point to the event and the Coushy horizons, as
the Riessner-Nordstorm metric \cite{Poisson}. Hence, we refer to
them as quasi event and quasi Coushy horizons. The case with $Q=0$,
$M=0$ and $\Lambda>0$ has attractive properties. Because in the low
expansion regime $(a(t)\simeq c)$, one can rewrite this case as
\begin{eqnarray}
ds^2\approx -(1-\frac{\Lambda}{3}\eta^2)dt^2+\frac{d\eta^2}
{(1-\frac{\Lambda}{3}\eta^2)}+\eta^2 d\Omega^2.
\end{eqnarray}
This is nothing but the De-Sitter spacetime with cosmological
constant $\Lambda$, which points to the current acceleration era.

\subsection*{Horizons and temperature}

Different $f(t,r)$ yield apparent horizons with different locations,
and one can use Eqs.~(\ref{aph1}) and~(\ref{sg1}) in order to find
the location and the temperature of apparent horizon. For every
$f(t,r)$, using the slow expansion regime, we get:
\begin{eqnarray}\label{Metricf}
ds^2\approx -f(\eta)dt^2+\frac{d\eta^2}{f(\eta)}+\eta^2 d\Omega^2.
\end{eqnarray}
Now, the location of horizons and their surface gravity can be
evaluated by using Eq.~(\ref{SG}). Their temperature is
approximately equal to Eq.~(\ref{Temp2}), or briefly:
\begin{eqnarray}
T_i \simeq \frac{f^{\prime}(\eta)}{4 \pi}|_{\eta_{hi}},
\end{eqnarray}
where $(^\prime)$ is derivative with respect to radii $\eta$ and
$\eta_{hi}$ is the radii of i$^{\textmd{th}}$ horizon.
\section{Conclusions \label{Conclusions}}
We considered the conformal form of the special group of the
non-static spherically symmetric metrics, where it was assumed that
the time dependence of the conformal factor is like as the FRW's. We
saw that the conformal BHs can be classified as a special subgroup
of these metrics. In order to derive the new solutions of the
Einstein equations, we have imposed perfect fluid concept and used
slow expansion approximation which helps us to clarify the physical
meaning of the parameters of metric. Since the Einstein tensor is
diagonal, there is no energy accretion and thus the backreaction
effect is zero. This imply that the energy (mass) should be constant
in our solutions. These new solutions have similarities with earlier
metrics that have been presented by others \cite{mcvittie,Gao,Gao0}.
A related metric which is similar to the special class of our
solutions was introduced by mcVittie \cite{mcvittie,Gao}. These
similarities are explicit in the flat case (temperature and entropy
at the redshift singularity), but the differences will be more clear
in the non-flat case ($k\neq0$), and we pointed to the one of them,
when we debate the redshift. In addition and in the flat case, we
tried to clear the some of differences between our solution and the
mcVittie's. We did it by pointing to the behavior of the redshift
singularity in the various coordinates, the mass notion, and
thermodynamics. Meanwhile, when our slow expansion approximation is
broken then there is no horizon for our solutions. Indeed, these
objects can be classified as naked singularities which can be
considered as alternatives for BHs \cite{virb,virb1}.

For the our solutions and similar with earlier works
\cite{mcvittie,Gao,Gao0}, the co-moving radiuses of the redshift
singularities are decreased by the expansion of universe. Also,
unlike the previous works \cite{mcvittie,Gao,Gao0}, the redshift
singularities in our solutions are independent of the background
curvature. By considering the slow expansion approximation, we were
able to find out BH's like behavior of these singularities. We
pointed to these objects and their surfaces as quasi BHs and the
quasi horizons, respectively. In continue, we introduced the
apparent horizon for our spacetime which should be evaluated by
considering the FRW background.

In order to compare the mcVittie's solution with our mcVittie's like
solution, we have used the three existing definitions of mass
including the Komar mass, the Misner-Sharp mass ($M_{MS}$) and the
Gong-Wang mass ($M_{GW}$). We saw that the notion of the Komar mass
of quasi BH differs from the mcVittie's solution. Also, in our
spacetime, we showed that the $M_{MS}$ and $M_{GW}$ masses yield the
same result on the apparent horizon and cover the FRW's result in
the limiting situations. In addition, using the slow expansion
approximation, we evaluated $M_{MS}$ and $M_{GW}$ on the quasi event
horizon of our mcVittie's like solution, which leads to the same
result as the Komar mass. In addition, we should express that, the
same as the mcVittie spacetime, the energy conditions are not
satisfied near the quasi horizon.

In addition, we have proved that, unlike the mcVittie's solution,
the first law of thermodynamics may be satisfied on the quasi event
horizon of our mcVittie's like solution, if we use the Komar mass or
either $M_{MS}$ or $M_{GW}$ as the confined mass and consider the
slow expansion approximation. This result is consistent with
previous studies about the conformal BHs \cite{Fara}, which shows
that the thermodynamics of our solutions is similar to the conformal
BHs'. In order to clarify the mass notion, we think that the full
analysis of the Hawking-Hayward mass for our solution is needed,
which is out of the scope of this letter and should be considered as
another work, but our resolution makes this feeling that the
predictions by either the slow expansion approximation or using the
suitable coordinates for describing the metric for mass, may be in
line with the Hawking-Hayward definition of energy, and have
reasonable accordance with the Komar, $M_{MS}$ and $M_{GW}$ masses
of our solutions. Indeed, this final remark can be supported by the
thermodynamics considerations and the no energy accretion condition
($G_{tr}=0$). Moreover, we think that, in dynamic spacetimes, the
thermodynamic considerations along as the slow time varying
approximation can help us to get the reasonable assumptions for
energy and thus mass. Finally, we saw that the first law of
thermodynamics will be approximately valid in the mcVittie's and our
solutions if we use the Hawking-Hayward definition of the mass in
the mcVittie spacetime and the Komar mass as the physical mass in
our solution, respectively. In continue, the more general solutions
such as the charged quasi BHs and the some of their properties have
been addressed.

Results obtained in this paper may help achieving a better
understanding of black holes in a dynamical background. From a
phenomenological point of view, this issue is important since after
all, any local astrophysical object lives in an expanding
cosmological background. Finally, we tried to explore the concepts
of mass, entropy and temperature in a dynamic spacetime.
%%%%%%%%%%%%%%%%%%%%%%%%%%%%%%%%%%%%%%%%%%%%%%%%%%%%%%%%%%%%%%%%%%%%%%%%%%%
\section{Acknowledgments}
We are grateful to referee for appreciable comments which led to
sensible improvements in this manuscript. This work has been
supported financially by Research Institute for Astronomy \&
Astrophysics of Maragha (RIAAM).
%%%%%%%%%%%%%%%%%%%%%%%%

\end{document}